\documentclass{aa}

\usepackage{epsfig}
\usepackage{graphics}

\newcommand{\be}{\begin{equation}}
\newcommand{\ee}{\end{equation}}

\begin{document}

\title{Eddington-Malmquist bias in a cosmological context}
\author{P. Teerikorpi}
\institute{Tuorla Observatory, Department of Physics and Astronomy,
 University of Turku, FIN-21500 Piikki\"{o}, Finland}
\date{Received / Accepted}

\abstract
{}
{
In 1914, Eddington derived a formula for the difference between the mean absolute magnitudes of stars "in space" or gathered "from the sky". Malmquist (1920) derived a general relation for this difference in Euclidean space.
Here we study this statistical bias in  cosmology, clarifying and expanding previous work.
}
{We derived the Malmquist relation within a general cosmological framework, including Friedmann's model,  analogously to the way Malmquist showed in 1936 that his formula is also valid in the presence of extinction in Euclidean space. We
also discuss some conceptual aspects that explain the wide scope of the bias relation.
 }
{The Malmquist formula for the intrinsic difference  $\langle M \rangle_m - M_0 = - \sigma_M^2 \frac{ \mathrm{d}\ln{ a(m)}}{\mathrm{d}m}$ is also valid for observations made in an expanding Friedmann universe. This is holds true for bolometric
and finite-band magnitudes when $a(m)$ refers to the distribution of observed 
(uncorrected for K-effect or $z$-dependent extinction) 
apparent magnitudes.
}
{}


\authorrunning{P. Teerikorpi}
\titlerunning{Eddington-Malmquist bias in a  cosmological context}

\maketitle

\section{Introduction}

Although not an astronomer, but a philosophical and cosmological thinker, Giordano Bruno already understood that our fixed position in the Universe leads to problems for the observer. When he discussed his vision of the cosmos filled with stars and planetary systems, he pointed out some reasons why we cannot see all those planets: 1) they may be too faint, 2) they may be too far away, and 3) they may reflect the light of their central star only
poorly (in
the third dialog of his book {\it Of the infinite universe
and worlds,} 1584). 

Thus Bruno was faintly aware of some selection effects that make the life (or at least the task) of an astronomer difficult. Before the telescopic era, Bruno had in mind visual observations, but similar problems affect observations made with instruments.   The fact that celestial bodies have a range of luminosities and are scattered in space leads to interesting effects that the observer should take into account.

\subsection{Eddington's contribution}

In his book {\it Stellar movements and the structure of the universe,
} 
Eddington (\cite{eddington14}) discusses in the chapter "Phenomena associated with spectral type" among other topics the distribution of stars in the Hertzsprung-Russel (HR) diagram (absolute magnitude vs. spectral type) that had recently been introduced by Hertzsprung and Russell. They had proposed the existence of "giant" and "dwarf" stars. Eddington begins by considering what happens when one chooses Type A and Type M stars "at random out of the stars in space", and then he says:

\vspace{0.2cm}
{\it We say intentionally "out of the stars in space", because, for example, the stars visible to the naked eye are a very special selection by no means representative of the true distribution of the stars.}

\vspace{0.2cm}
When discussing the reality of the dwarfs and giants division, Eddington then ponders the possibility of a selection effect. As the stars for which the parallaxes had been derived had been chosen either for brightness or for nearness (large proper motion), he asks if the two groups might result "from the double mode of selection, without implying any real division in the intrinsic luminosities".

Then he shows in a few strokes that for a normal distribution of absolute magnitudes $M$ with the mean $ M_0$ and the dispersion $1/(\sqrt{2} k)$ (his notation), the frequency of $M$ among stars in a magnitude-limited sample is a normal distribution with the same dispersion, but a brighter mean value $M_0 - 0.69/k^2$. With this formula at hand, Eddington shows that to explain in this way the wide gap of $11$ mag between  giants and dwarfs for the M spectral type would require a dispersion of almost 3 mag, while Russell had derived a mean value of $1.14$ mag for all spectral types. Thus he concludes that the selection effect, which he had just discovered, cannot explain dwarfs and giants. 

In the current notation Eddington's formula is written
\begin{equation}
\langle M \rangle = M_0 - 1.382 \sigma_M^2  ,
\end{equation}
where $\langle M  \rangle$ is the mean absolute magnitude when a stellar class with a Gaussian luminosity function (LF) ($M_0 =$ the mean absolute magnitude, $\sigma_M=$ the dispersion) is sampled from the sky as a magnitude-limited sample. Such a class may be called a Gaussian standard candle.  
 
\subsection{General Malmquist formula}

Equation (1) is valid for a homogeneous spatial distribution. It is a special case of the more general formula derived by Malmquist (\cite{malmquist20}, \cite{malmquist22}). He investigated how the luminosity function (LF) of A-type stars may be derived from their distances (in fact, proper motions), provided that it is Gaussian and one knows the distribution $a(m)$ of apparent magnitudes up to a limiting magnitude. One result was the Malmquist formula for the mean value of $M$ of a sample gathered through the apparent magnitude "window" $m \pm \frac{1}{2}\mathrm{d}m$:
\be
\langle M \rangle_m = M_0 - \sigma_M^2 \frac{ \mathrm{d}\ln{ a(m)} }{\mathrm{d}m}
.\ee
The $a(m)$ term in Eq.(2) depends on the spatial distribution of stars. It has a simple constant form, when the number density reads  $r^{-\alpha}$, where $r$ is the distance:
$\langle M \rangle_m  = M_0 - (3 - \alpha) 0.461 \sigma_M^2$.
With $\alpha = 0$, the Eddington Eq. (1) is obtained.
A strongly thinning density with $\alpha = 3$  is required for no bias, $\langle M \rangle_m = M_0$. Then the large volume at large distances, which contributes high-luminosity stars to the sample, is fully compensated for by the lower number density of stars.

\subsection{About terminology}

The Eddington-Malmquist (or Malmquist) bias refers to the major difference in sampling luminous objects "from space" versus "from sky".
The Malmquist relation is the general  Eq. (2), while the Eddington formula is the special case, Eq. (1).

Butkevich et al. (\cite{butkevich05}) termed the bias in Eq.(2) differential, while integral bias was used to denote Malmquist's other formula,
\be
\langle M \rangle_{\mathrm{int}} = M_0 - \sigma_M^2 \frac{\mathrm{d} \ln A(m_{\mathrm{lim}})}{\mathrm{d}m_{\mathrm{lim}}}
,\ee
where $\langle M \rangle_{\mathrm{int}}$ is the mean for the whole magnitude-limited sample
and $A(m)$ is the cumulative distribution up to the magnitude limit $m_{\mathrm{lim}}$. 

When standard candle data are inspected as $\langle M \rangle$ vs. $m$ or $\langle M \rangle$ vs. $r$ ($r=$ distance), respective biases of Type 1 and Type 2
appear (as reviewed by Teerikorpi \cite{teerikorpi97}, especially Table I therein), which were also called classical and distance-dependent by Sandage (\cite{sandage94}).

Type 1 relates to the bias treated here, that is,
how $\langle M \rangle_m$ differs from $M_0$.
Type 2 refers to the magnitude cut-off effect when for instance
a Hubble diagram is inspected as $m$ versus $\log z$. Often the Type  2 aspect is also called, a little misleadingly, the Malmquist bias.

Another parameter is the Eddington bias, which denotes the influence of random measurement errors
on derived distribution functions (Eddington \cite{eddington13}, \cite{eddington40}). The Eddington bias was discussed by Teerikorpi (\cite{teerikorpi04}), who also considered how it works in concert with the Malmquist bias, Eq.(2).

We derive in Sect. 2 the Malmquist relation in a general
cosmological context, first using the bolometric magnitude and then for a finite-band magnitude.  
In Sect. 3, the result is illustrated and compared
with our earlier studies. 
Section 4 contains concluding remarks.  

\section{Malmquist equation in a cosmological context}

We have previously discussed cosmological Malmquist bias in the Hubble diagram at high redshifts (Teerikorpi \cite{teerikorpi98}, \cite{teerikorpi03}; or T98, T03). This was made by calculating
the behaviour of the average $\langle \log z \rangle_m$ for a Gaussian standard candle, taking into account the different foreground
and background volumes as given by Friedmann models. The Malmquist formula was not directly considered.

An early work on the Malmquist bias in cosmology was made by Bigot \& Triay
(\cite{bigot90}), kindly communicated
by them to us after the paper T98 was published. The present discussion should  facilitate access to their technical treatment,
where one result is the Malmquist integral relation, Eq.(3), and where they conclude that the constant correction
(Eq. (1)) is no longer valid for distant objects.

Here we derive the general formula using the cosmological route, but analogously to the way Malmquist (\cite{malmquist36}) remarkably showed that Eq.(2) is valid not only in
transparent Euclidean space, but  also in the presence of interstellar extinction. 
We first assume fully transparent space and start with the differential bias (Eq. (2)), whose derivation illustrates well
the classical and cosmological aspects of the bias and from which it is easy to derive the integral bias.

\subsection{Bolometric magnitude}

We begin with the necessary formulae using the bolometric magnitude.
Instead of the classical distance, we use the redshift $z$ as the parameter indicating the distance.
Then the observed apparent (bolometric) magnitude $m$ is related to the absolute (bolometric) magnitude $M$ as $M = m - \mu (z)$,
where $\mu (z)$ is the Friedmann model-dependent distance modulus of an object at redshift $z$. \footnote{One may write $\mu (z) = 5 \log {r_{\mathrm lum}} (z)/10{\mathrm pc}$.  The luminosity distance $r_{\mathrm lum} (z)$ is obtained using
the well-known Mattig equation and its generalizations (e.g., Baryshev \& Teerikorpi
\cite{baryshev12}), once the values of the Friedmann model parameters are fixed.}

We consider a class of objects with a Gaussian LF $\Phi (M)$ for the bolometric magnitudes. Then the number
of objects in the sky observed with the apparent magnitude $m \pm \frac{1}{2} dm$ (differential counts) may be obtained using the analogue of
the equation of  von Seeliger (\cite{seeliger98}), now summing over the redshift:
\be
a(m)\mathrm{d}m =  \frac{\omega}{4\pi}
\int_0^{\infty}\Phi (m - \mu (z))\rho (z) \frac{\mathrm{d}V}{\mathrm{d}z} \mathrm{d}z \mathrm{d}m 
,\ee
where $\rho (z)$ gives the co-moving spatial number density of objects, possibly varying as a function of redshift, $V(z)$ is the co-moving volume up to redshift $z$, and $\omega$ is the solid angle covered by the region under survey ($\frac{\mathrm{d}V}{\mathrm{d}z}$ is the co-moving volume derivative).

By derivation, one obtains another needed expression
\be
\frac{\mathrm{d} a(m)}{\mathrm{d}m} = \frac{\omega}{4\pi}  
 \int_0^{\infty}\frac{\mathrm{d} \Phi (m - \mu (z))}{\mathrm{d}m} \rho (z)\frac{\mathrm{d}V}{\mathrm{d}z} \mathrm{d}z
.\ee
Using Eq.(4), the average value of the absolute magnitude of the objects observed at $m \pm
\frac{1}{2} \mathrm{d}m$ reads
\be
\langle M \rangle_m a(m) = \frac{\omega}{4\pi}   
\int_0^{\infty}(m-\mu (z)) \Phi (m - \mu (z))\rho (z)\frac{\mathrm{d}V}{\mathrm{d}z} \mathrm{d}z
.\ee
Inserting the Gaussian LF, Eq. (5) becomes
\begin{eqnarray}
\frac{\mathrm{d} a(m)}{\mathrm{d}m} =  - \frac{\omega}{4\pi}\int_0^{\infty}\frac{m-\mu (z) - M_0}
{\sigma_M^2}\Phi (m - \mu (z))\rho (z)
\frac{\mathrm{d}V}{\mathrm{d}z} \mathrm{d}z \nonumber \\
= -\frac{1}{\sigma_M^2}(\langle M \rangle_m a(m) - M_0 a(m))
,\end{eqnarray}
from which one finally obtains
\be
\langle M_{\mathrm{b}} \rangle_{m\mathrm{b}} = M_{0\mathrm{b}} - \sigma_M^2 \frac{\mathrm{d} \ln {a(m_{\mathrm{b}})}}{\mathrm{d}m_{\mathrm{b}}}
,\ee
where b means that the magnitudes are bolometric.
 Equation (8) is the same as the classical Malmquist formula. Below we discuss
some further aspects of the result.

\subsection{K-correction $K(z)$ and  extinction $E(z)$}

The above derivation, with the bolometric magnitude in cosmology, 
corresponds to the case of transparent Euclidean space in
Malmquist's original study, where the apparent magnitude could be 
bolometric or finite-band (no redshift).
  
In practice, a finite-band magnitude $m_i$ is
measured that gives rise to a redshift-dependent K-effect $K_i (z)$. In that case, one
replaces $M = m - \mu (z)$ by $M_{i, \mathrm{c}} = m_i - K_i (z) - \mu (z), $ and the end result is similar to Eq.(8), now for the K-corrected $M_i$-magnitude $M_{i,c}$ and the observed $m_i$ magnitude:
\be
\langle M_{i, \mathrm{c}} \rangle_{m_{\mathrm{i}}} = M_{i0} - \sigma_M^2 \frac{\mathrm{d} \ln {a(m_{\mathrm{i}})}}{\mathrm{d}m_{\mathrm{i}}}
.\ee
We emphasize a subtlety in  the magnitudes in Eq.(9). The difference $\langle M_{i, \mathrm{c}} \rangle_{m_{\mathrm{i}}} - M_{i0}$ indicates how much the mean value of the  {\it intrinsic} (K-corrected) absolute magnitude of the objects at the {\it observed} apperent
magnitude $m_{\mathrm{i}}$ differs from the actual mean $M_{i0}$ of
the Gaussian LF. In the right-side expression, 
the distribution  $a(m_{\mathrm{i}})$ is that of the observed (uncorrected)
apparent magnitude.

Adding a $z$-dependent
extinction $E(z)$ to the model ($M_{i, \mathrm{c}} = m_i - K_i (z)  - E(z)- \mu (z) $) leads to the same result.

 The symbol $m$ may designate either a bolometric or (as in practice) a finite-band
magnitude in the remaining text.

\subsection{Integral relation}

In the integral bias the relevant variable is the limiting magnitude $m_{{\mathrm lim}}$ up to which the sample is complete in the
inspected region of the sky (in the derivation of the differential bias it is not required that the sample is complete in this sense). Then up to $m_{{\mathrm lim}}$, the number of objects is
\be
A(m_{\mathrm{lim}}) =   \int_{-\infty}^{m_{\mathrm{lim}}} a(m) \mathrm{d}m
,\ee
and the mean absolute magnitude for the whole sample is
\be
\langle M \rangle_{\mathrm{int}} A(m_{\mathrm{lim}}) = 
\int_{-\infty}^{m_{\mathrm{lim}}} \langle M \rangle_{m} a(m) \mathrm{d}m  
.\ee
From what was discussed above, we know what $\langle M \rangle_{m}$ is (i.e., Eq.(9)) both classically and cosmologically for
a Gaussian LF, and inserting this into Eq.(11) results in
\be
\langle M \rangle_{\mathrm{int}} A(m_{\mathrm{lim}}) = M_0 A(m_{\mathrm{lim}}) - \sigma_M^2
 \frac{\mathrm{d} A(m_{\mathrm{lim}})}{\mathrm{d}m_{\mathrm{lim}}}
,\ee
from which follows Eq.(3).

It is interesting to note that the Malmquist differential and integral relations also apply to the extreme spatial distribution $z =$ constant.
This is considered in Appendix A. 

\subsection{Malmquist relation via convolution}

The Malmquist relation can also be considered via convolution.
Namely, the distribution $a(m)$ results from a convolution of a Gaussian
function $\Phi$ and a function $F$ when (e.g., in the cosmological
context) one considers the distance modulus $\mu$ as a variable instead of $z$.
Then
\be
F(\mu) = \frac{\omega}{4\pi} \rho (z (\mu)) \frac{\mathrm{d}V}{\mathrm{d}z}(z (\mu))
\frac{\mathrm{d}z (\mu)}{\mathrm{d}\mu} , 
\ee
\be
a(m) =  \int_{-\infty}^{\infty}
\Phi (m - \mu) F(\mu){\mathrm{d}\mu} = (\Phi \star F) (m) ,
\, \, \, \mathrm{and}
\ee
\be
\Delta M_m =
 - \sigma^2 \frac{\mathrm{d} \ln {(\Phi \star F) (m)}}{\mathrm{d}m}
.\ee

Here we have the essential mathematical reason for the wide scope
of the Malmquist relation because it is based on the simple Gaussian convolution of the function
$F,$  which carries 
all cosmological factors 
(geometry, luminosity distance, possible number density evolution),
and the distance modulus $\mu$ may be viewed as a dummy integration variable.
With the K-effect  (Sect.2.2),
the new variable is constructed 
from $\mu (z)+ K(z)$, which is normally a monotonically increasing function. \footnote{For
a class of identical objects ($M = M_0$), $\Phi$ $\propto$
the Dirac $\delta$-function, and the convolution results in $a(m) \propto F(m-M_0)$.
} 

\section{Discussion}
 
To illustrate the result in a concrete way, it is instructive to make numerical calculations of the left and right sides of the Malmquist equation in
Friedmann space. 

\subsection{Illustrations of the result}

For example, consider a class of Gaussian standard candles with
$M_0 = -23$ mag and $\sigma_M = 0.4$ mag. Bolometric magnitudes
are first assumed and the Einstein-de Sitter model with $H_0 = 50$ km/s/Mpc
is used (the exact value of $H_0$ is not relevant). Then the distance
modulus is $\mu = 5 \log (1+z - (1+z)^{1/2}) + 45.4$ and the co-moving
volume derivative is $\frac{dV}{dz} \propto ( (1+z)^{1/2} -1)^2 /(1+z)^{5/2}$.

Using these relations in Eqs. (4) and (6), with $\rho (z)=$ constant (no number
evolution), we calculate the logarithmic distribution of apparent magnitudes $a(m)$
as shown in Fig.1 (an arbitrary zero-point) and the average $M$ at different observed $m$, or 
$\langle M \rangle_m$. The slope of $\log a(m)$ is indicated for a few
apparent magnitudes. Note the expected classical slope $0.6$ at bright
magnitudes (low redshifts).

Calculation shows that the expression 
$M_0 - (\mathrm{d} \log {a(m)}/\mathrm{d}m)/0.6 \times 1.382 \sigma_M^2$
indeed reproduces the numerically calculated $\langle M \rangle_m$ (see the
upper part of Fig.1).
 
Of course, this agreement is just as expected from
the derived Malmquist relation. However,  in presenting Fig.
1, we wish to underline several aspects. In the distribution $a(m)$, $m$ is the {\it observed, uncorrected}
apparent magnitude. The difference $\langle M \rangle_m -M_0$ for intrinsic absolute magnitudes
can be derived without detailed information on the Friedmann model in question.  This is also true for some number 
evolution ($\rho (z)$). As the slope of $\log a(m)$
decreases starting from $0.6$, the Malmquist bias decreases
for this  model as well.

Figure 2 presents similar calculations, but now assuming finite-band apparent
magnitudes subject to a K-effect. For simple illustration, the K-correction is
taken to be $K(z) = 2.5z$ mag, roughly like for elliptical
galaxies in optical wavebands (Coleman et  al. 1980), making them apparently fainter
than would be the bolometric expectation at increasing redshifts. 
Again, here $a(m)$ is the raw distribution of observed  magnitudes,
without the K-correction. 

\begin{figure}
\epsfig{file=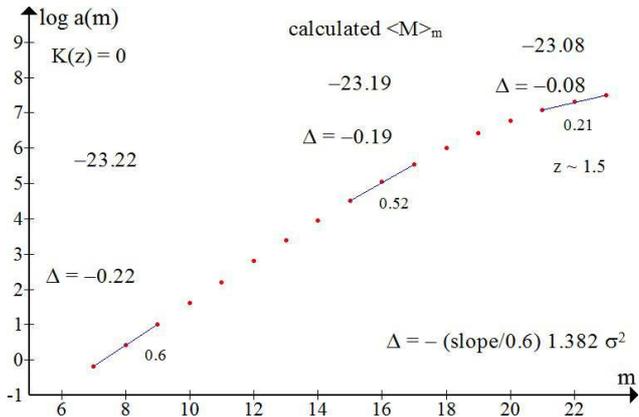 , angle=0, width=8.3cm}
\caption{$\log a(m)$ vs. bolometric magnitude $m$ (Eq.(4)) for a standard
candle with $M_0 = - 23.0$ and $\sigma_M = 0.4$ mag (the E-deS model, see text). The upper figures
above the curve are the mean values  $\langle M \rangle_{m}$ calculated from Eq. (6) for $m =$ 8, 16 and 22 mag,
respectively. The lower figures ($\Delta$) are the Malmquist bias values as calculated from the slopes shown below the curve.    
}
\end{figure}

\begin{figure}
\epsfig{file=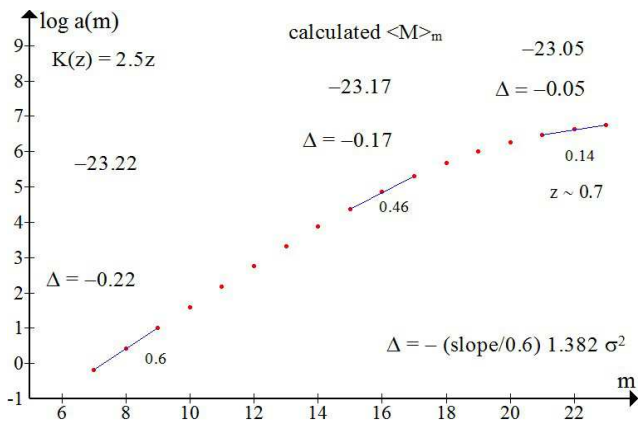 , angle=0, width=8.3cm}
\caption{$\log a(m)$ vs. finite-band apparent magnitude $m$ for a Gaussian standard
candle. Here the K-effect needed in the calculations of $a(m)$ and the average absolute magnitude
$\langle M \rangle_{m}$
corresponds to the correction $K(z) = 2.5z$. Other parameters have the same values as in Fig.1.
This and Fig.1 also show roughly where the redshift distribution peaks around $m=22$ mag.     
}
\end{figure}

\subsection{Comparison with the bias analysis in T98} 

We have explained what happens to the Malmquist bias in cosmology in terms
of the luminosity distance and the corresponding comoving volume (T98).
As the cosmological distances and volumes are related in a way
different from the classical distances,
the bias is generally not constant in Friedmann models, but depends on apparent magnitude. 

In the cited analysis, instead of $\langle M \rangle_{m}$, we calculated
the quantity $\langle \log z \rangle_m$, which is directly suitable to analyse the Hubble $\log z $ vs. $m$ diagram.  A uniform spatial distribution was assumed with no comoving number evolution.

We conclude 
that the  curves of Mattig (\cite{mattig58}) for Friedmann models
need to be corrected for a
non-constant Malmquist bias in the $\log z$ vs. $m$ Hubble diagram (T98; also Bigot \& Triay \cite{bigot90}).

It was found that at bright $m$ (low $z$) the Malmquist
shift is close to classical, as expected. Then it generally decreases (in absolute value) towards fainter magnitudes.
The same  can now be seen from the Malmquist relation.

 When changing the  model for example by adding the
cosmological constant $\Lambda$,
one simply asks how the slope of the counts $a(m)$ changes. 
 While the comoving volume derivative $V'(z)$ 
becomes steeper with increasing $z$ when a positive $\Lambda$ is added, (tending to increase the slope), the luminosity
distance also increases (diminishing the slope).

The rapidly increasing volume is more important; it results in steeper $a(m)$, which means a larger Malmquist bias, that is, closer to the classical one.
For example, referring to Fig.1, at $m = 22 $ mag the pure flat $\Lambda$ model, $\Omega_{\Lambda} = 1$, would
lead  to the bias $-0.15$ instead of $-0.08$ for the E-deS flat model.
Thus the $\Lambda$-model leads to a weaker $m$-dependence of the bias, as also derived in T98 and T03.   

For the K-effect, it was pointed out in T98 that if $K(z)$ increases with $z$ so that
the objects become fainter quicker than when they are only due to the bolometric factor, then the backside volume effectively
decreases and the trend in the Malmquist bias away from the classical case increases, as also seen here in Figs. 1 and 2. The K-effect can be important and increase the deviation of the bias from the classical constant value.

\subsection{Practical note}

That we can derive the difference of the intrinsic magnitudes $\langle M \rangle_{m}$ and $M_0$
 in principle  from minimal information ($\sigma_M$) and raw data ($a(m)$)   
does not mean that we may generally forget factors such
as the K-correction when applying this result. 

For example, we consider a Gaussian standard candle in a test
of the Friedmann model, assuming that we have been able to determine
$a(m) $ and know $\sigma_M$. 

At the observed $m$ the average
$\langle M \rangle_{m}$
is predicted to be $M_0 - \sigma_M^2 \frac{\mathrm{d} \ln {a(m)}}{\mathrm{d}m}$.
This value is compared with the average
$\langle M\rangle_{m, \mathrm{data}}$ derived from the K-corrected apparent magnitudes of
the objects at the observed (uncorrected) magnitude $m$. Each object has a known redshift, so one may calculate for each its K-corrected absolute magnitude
for a given Friedmann model. Therefore the test requires knowing
$K(z)$ and the expression for the luminosity distance.

Referring to Fig. 2, one might have derived the slope 0.46 at $m = 16$ and hence the bias $-0.17$ mag.
Then the Friedmann model is correct, which gives $\langle M \rangle_{16, \mathrm{data}} =  -23.17$ mag, as derived from the K-corrected apparent magnitudes of the objects at the uncorrected $m = 16$ mag. 

With all the data, one requires that the difference 
\be
\Delta M_m= \langle M \rangle_{m, \mathrm{data}} - \big(M_0 - \sigma_M^2 \frac{\mathrm{d} \ln {a(m)}}{\mathrm{d}m}\big)
\ee
does not depend on $m$.
 In addition, $\langle \Delta M_m \rangle \approx 0$ for
the correct Hubble constant and $M_0$. 
\footnote{The same is valid in stellar statistics if the Malmquist relation is to be used to derive the mean absolute magnitude of a stellar class. 
$\langle M \rangle_\mathrm{data} - M_0$ is derived from the histogram of the apparent
magnitudes. Then to infer $M_0$, 
$\langle M \rangle_\mathrm{data}$ must be computed, which requires distances and extinctions for each sample star.} 

\section{Concluding remarks}

Malmquist (\cite{malmquist36}) reported that his relation is also valid when light extinction is added
to the static Euclidean space he considered. His study inspired the present work, which
shows the scope of the Malmquist relation from classical situations to 
Friedmann cosmological models.

The cosmological factors (luminosity distance, comoving volume derivative, and number
evolution) are all reflected in the slope of the (log) counts vs. observed apparent
magnitude. The K-effect for finite-band magnitude and also possible $z$-dependent extinction
 are automatically included in the right side of the Malmquist  relation.  

We emphasized conceptual aspects of the Malmquist relation in view
of its important role in stellar statistics.  
However, prospects for its practical use for high-luminosity objects 
in extragalactic astronomy are not so immediate.

First, a constant Gaussian LF is rare or absent for objects found at low and high redshifts. 
 Second, it requires many data to dermine $\frac{\mathrm{d} \ln {a(m)}}{\mathrm{d}m}$ with good accuracy.
 In addition, the Type 1 Malmquist analysis is too simplistic if the objects
are not detected on the basis of non-variable brightness, but there is a chain of measurement and
luminosity inference as for Ia supernovae.  

The Type 2 approach is often applied in
luminosity-bias analysis
in extragalactic astronomy, where the redshift offers a good relative distance
indicator. This concerns the classical determination of the Hubble constant as well as the detection of universal acceleration from the Hubble diagram
of Ia SNe (e.g., Perrett et al. \cite{perrett10}).
Bayesian approaches are also currently used when considering
general LFs and inferring cosmological model parameters
(e.g., March et al. \cite{march11}).

The Malmquist correction may be viewed as a Bayesian approach to the problem of deriving $\langle M \rangle_m$ (and perhaps using it in distance estimation) when the standard candle has a known Gaussian LF as prior, while the Type 2 approach corresponds to the frequentist view of probability (Hendry \& Simmons 1994).
In the Gaussian case, the mean value of the posterior distribution $P(M \mid m)$ in the Bayesian formula
\be
P(M \mid m) P(m) = P(m \mid M) p(M)
\ee
can be solved directly, essentially following Sect.2.

\section*{Acknowledgements}
I thank Alexei Butkevich for help in providing me with the relevant chapter in Eddington's rare 1914 book and for discussions on different aspects of the Malmquist bias.
Good remarks by the  referee are also acknowledged.

\appendix

\section{Case of a cluster ($z=$ const.)}
In the integrations of Sect.2 (e.g., in Eq. 4) integration can
be restricted to a finite $z$ range
(where the objects in question exist). The density law $\rho (z)$
takes care of this. 

\begin{figure}
\epsfig{file=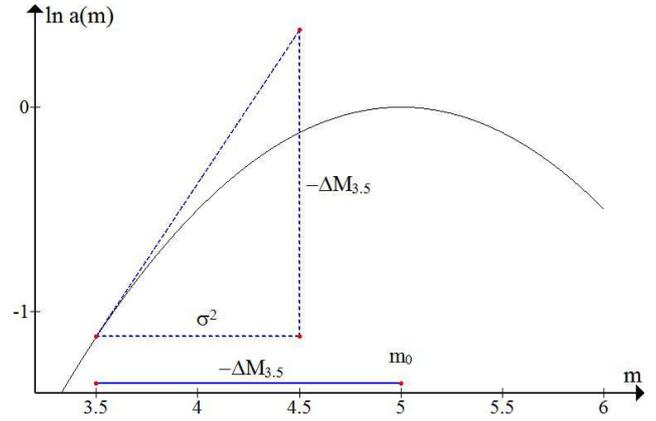 , angle=0, width=8.3cm}
\caption{Differential Malmquist relation as applied to the case
of a cluster at distance modulus $\mu = 10$. The LF is Gaussian
with $M_0 = - 5$ and $\sigma = 1$ mag. The slope of
the $\ln a(m)$ curve at any $m$, multiplied by $\sigma^2$, gives the difference $m_0 -m$. 
}
\end{figure}

Equation (9) also applies for a $\delta$-function-like $\rho (z)$, for instance, for a cluster at $z=z_0$.
Then $a(m)$ is proportional to a
Gaussian function with the mean $m_0 = M_0 + \mu (z_0)$ and is observed up to the sample limit $m_{\mathrm{lim}}$. The second term in the right side of Eq.(9) for any $m < m_{\mathrm{lim}}$ becomes after derivation
$m - M_0 - \mu (z_0) = M_m - M_0$. This is equal to
$\langle M \rangle_m -M_0$ in this case and gives the difference
between $m$ and the magnitude $m_0$
corresponding to the maximum of the LF at $M_0$ even if this
is not reached by the sample.  

The slope of the $\ln a(m)$ distribution
at any $m$ gives, via the Malmquist  Eq.(9), the absolute magnitude 
$M_m$ corresponding to the observed $m$ and hence the distance modulus
(see Fig. A.1 for a graphic representation).
For a good Gaussian LF the same result is obtained for any other $m$ in the
complete part of the sample. 

It is interesting to note that in the same year as Eddington published his
Eq. (1), Kapteyn (\cite{kapteyn14}) discussed the derivation of the distance of 
a star cluster with a Gaussian LF. He
considered the observational cut-off effect and
derived an integral equation that took into account the magnitude limit 
$m_{\mathrm{lim}}$
and contained the distance modulus as unknown. Essentially, in that method
the cluster is moved along the line of sight up to the distance modulus $\mu$ where the observed average apparent magnitude is equal to the value predicted from $\mu$, $m_{\mathrm{lim}}$, and the known Gaussian LF.  Of course, the end result
will be the same as in the approach where the slope of $\ln a(m)$
is used.

In fact, the formula (69) in the study by Kapteyn  is a special case of the Malmquist integral bias relation,
Eq.(3). Therefore both the differential and integral bias relations
by Malmquist are formally applicable to the extreme spatial distribution represented by $z=$ constant. 

{}

\end{document}